\documentstyle[epsfig]{aipproc}
\newcommand{\ppbar}{$p\overline{p}\ $}

\newcommand{\ebar} {\hbox{E\kern-0.5em\lower-0.1ex\hbox{/}}}
\begin{document}

\vspace*{-0.5cm}

\title{Prospects for Diffractive Physics \\
with the CDF Forward Detectors \\
at the Tevatron}
\author{Michele Gallinaro\thanks{Email: michgall@fnal.gov}$^,$\thanks{
Presented at ``LISHEP 2002 - Session C:Workshop on Diffractive Physics'', February 4-8 2002,
 Rio de Janeiro - RJ - Brazil.} }
\address{(CDF Collaboration) \\
 The Rockefeller University \\
 1230 York Avenue, New York, NY 10021 - USA } 

\maketitle

\begin{abstract}
The Forward Detector upgrade project at CDF
is designed to enhance the capabilities for studies of 
diffractive physics at the Tevatron during Run~II.
Studies of hard diffraction and very forward physics are some of the topics
that can be addressed in the next few years at the Tevatron.
The program for diffractive physics, including the detectors and 
their commissioning, is discussed here.
All the detectors have been installed and are presently collecting data.
\end{abstract}

\section{Introduction}
\begin{figure}[htb]
\begin{minipage}[thb]{1.\linewidth}
\epsfxsize=0.85
\textwidth
\centerline{\epsfbox{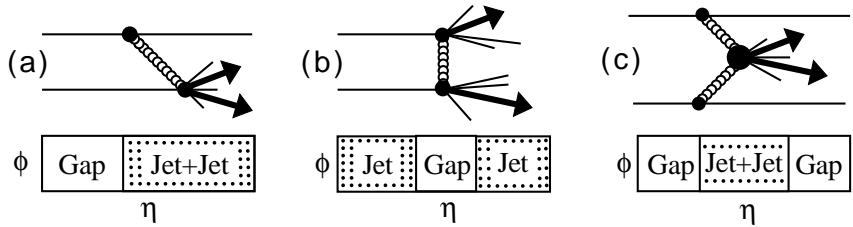}}
\caption{\label{fig:diffractive_diagrams}
Dijet production diagrams and event topologies for 
(a) single diffraction, (b) double diffraction, and (c) double Pomeron exchange.}
\end{minipage}
\end{figure}

\noindent
During Run~I, which started in 1992 and lasted until the end of 1995, the CDF experiment 
collected a large data sample which has been extensively studied. 
Considerable knowledge on diffractive physics phenomena was gained
using those data (see refs.~\cite{ken}--\cite{diff_w}).\\
Run~II at the Tevatron Collider started in the spring of 2001.
Protons and antiprotons are colliding at an energy of $\sqrt{s} = 1.96$~TeV.
In preparation for Run~II, improvements were made to the accelerator with the goal of achieving
instantaneous luminosities of ${\cal L}\approx 5\cdot 10^{32}$cm$^{-2}$sec$^{-1}$.
Although the size of the data sample delivered up to now is still well below 
the expectations, the Tevatron will hopefully soon reach the design luminosity goal.
Both the CDF and the D\O~ experiments underwent major upgrade projects to
improve their detector capabilities.
The Forward Detector upgrade project at CDF will enhance the sensitivity for 
hard diffraction and very forward physics.\\
The signature of a typical diffractive event in \ppbar collisions is characterized by a 
leading proton or anti-proton 
and/or a region at large pseudorapidity with no particles, also known 
as {\it gap} region (Fig.~\ref{fig:diffractive_diagrams}).
In order to detect such events, forward regions in pseudorapidity are extremely important.\\
\noindent
The Forward Detector upgrade project includes 
the {\it Roman Pot} (RP) fiber tracker detectors to detect leading antiprotons, 
a set of {\it Beam Shower Counters} (BSCs)
installed around the beam-pipe at three (four) locations along the $p$($\overline p$)
direction to tag rapidity gaps at 5.5$<|\eta |<$7.5, 
and two forward {\it MiniPlug} (MP) calorimeters to cover the pseudorapidity
region 3.6$<|\eta |<$5.1.\\
All of the above detectors have been installed and are now 
fully integrated with the rest of the CDF detector.

\section{Diffractive Physics during Run II}
\noindent
Diffractive physics can be studied at the Tevatron Collider
using the Forward Detectors.
The physics topics to be addressed include studies of soft and hard diffraction,
searches for centauros and disoriented chiral condensates, and forward jet
production. The latter can probe both small-{\it x} and large-{\it x} regions
of the proton structure function.

\subsection{Hard Single Diffraction}
\noindent
Hard diffraction processes are hadronic interactions which incorporate a high
transverse momentum partonic scattering while carrying the characteristic
signature of diffraction, namely leading $p$ or $\overline{p}$ 
particles and/or rapidity gap regions.\\
Hard Single Diffraction (SD) studies can be used to determine the structure of the
pomeron, as well as to determine the dependence of the cross section on the
recoil (anti)proton fractional momentum loss ($\xi$)
and on the four-momentum transfer squared ($t$). 
The wide $\xi$ coverage (down to $\xi_{\rm min}$=0.03) will allow the
measurement of the dijet cross
section as a function of $\xi$ and thereby the determination of the pomeron contribution
to the cross section and the factorization properties of hard diffraction as a
function of $\xi$.

\subsection{Double Diffraction}
\noindent
The $\Delta\eta$ dependence of the ratio of dijet events with a rapidity 
gap between jets to non-gap events probes the coupling of the
exchanged color-singlet object to quarks and gluons relative to the coupling of
the normal color-octet exchange.\\
The MP can be used to study {\it soft} Double Diffraction (DD) 
where both proton
and anti-proton dissociate. This process, which is characterized by a rapidity gap
between two clusters of particles in the regions of large (positive and negative)
pseudorapidity, has been studied in Run~I~\cite{diff_ddd}. The MP will allow
the measurement to be extended to larger gaps.
The ratio of DD to {\it minimum bias} events may shed light 
on the mechanism of the survival probability of the rapidity gap in events with jets.

\subsection{Double Pomeron Exchange}
\noindent
This is an area of great interest for hard diffraction processes.
In Run~I, only a small number of Double Pomeron Exchange (DPE) dijet events 
were observed using a combination of an
anti-proton tag on the west side and a rapidity gap tag on the opposite
(east) side~\cite{diff_dpe}.\\
During Run~II, the momentum of the pomeron associated with the gap can be measured 
by the segmented MP calorimeter (by measuring the center-of-mass energy of
the pomeron-pomeron system using calorimeter information).
Low mass exclusive DPE events (events with rapidity gaps on both sides) 
can be searched for by using both the Plug and the MP calorimeters.
Furthermore, the exclusive dijet and $b\overline{b}$ production cross sections 
can also be measured.
The latter is of much interest to diffractive Higgs
production~\cite{rostov}, as recent studies~\cite{royon} indicate that Higgs boson
and dijet production via DPE may be non-negligible at the Tevatron.

\subsection{Forward Jet Production}
\noindent
The Bjorken scaling variable ({\it x}) can be studied, both in the 
small-{\it x} and the large-{\it x} regions of the proton structure function, 
by measuring the cross section of events with two jets in the forward regions.
The dijet events probe the parton density and do not discriminate between
gluons and quarks. 
Small-{\it x} values can be studied using Same Side (SS) dijet events, while
large-{\it x} regions can be explored with both SS or opposite-side (OS) dijet events.
The large-{\it x} gluon density of the proton can be probed 
in events with a prompt (isolated) photon in the central region and a 
forward jet in the MP. 

\subsection{Centauros and Disoriented Chiral Condensates}
\noindent
The unique signature for Centauro or Disoriented Chiral Condensates (DCC) 
events is a high multiplicity of
relatively low $P_{T}$ clusters of particles depositing predominantly hadronic
or electromagnetic energy. Such particles have been searched for in
Run~IA data~\cite{centauro} with negative results. The fine segmentation of the
MP and the excellent position resolution, not only for leptons and photons but
also for hadrons, makes the MP an ideal detector for searching for
Centauro and DCC events.

\section{Detectors}
\noindent
Theories and expectations cannot be proven, right or wrong, without experiments. 
Experiments need detectors.
The detectors built and installed at
CDF during the preparation for Run~II with the aim of addressing the above topics
are described in the following sections.

\subsection{Roman Pot Fiber Tracker}
\noindent
The pomeron structure can be determined by using the kinematic variables in
diffractive dijet events, provided that the momentum of the pomeron is known.
During Run~I, the pomeron momentum was determined using the RP to measure the
momentum of the leading anti-proton.\\

\begin{figure}[thb]
\vspace*{0.5cm}
\begin{minipage}[]{.45\linewidth}
\epsfxsize=1.1\textwidth
\centerline{\epsfbox[30 440 555 740]{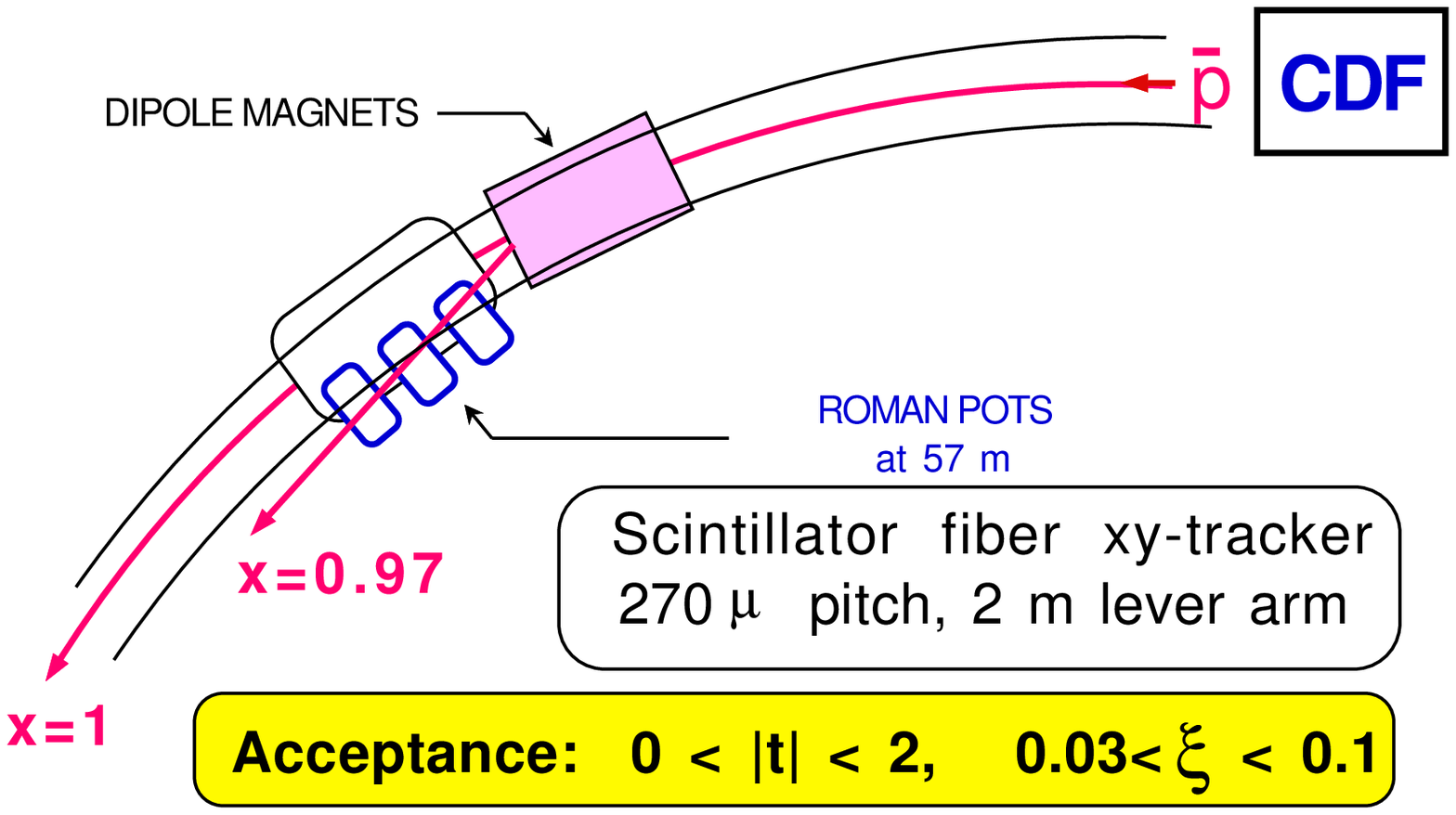}}
\vspace*{0.5cm}
\caption{\label{fig:rps_tevatron}
Schematic view of the Roman Pot spectrometer at the Tevatron.}
\end{minipage}
\hspace*{0.6cm}
\begin{minipage}[]{.45\linewidth}
\epsfxsize=1.1\textwidth
\centerline{\epsfbox{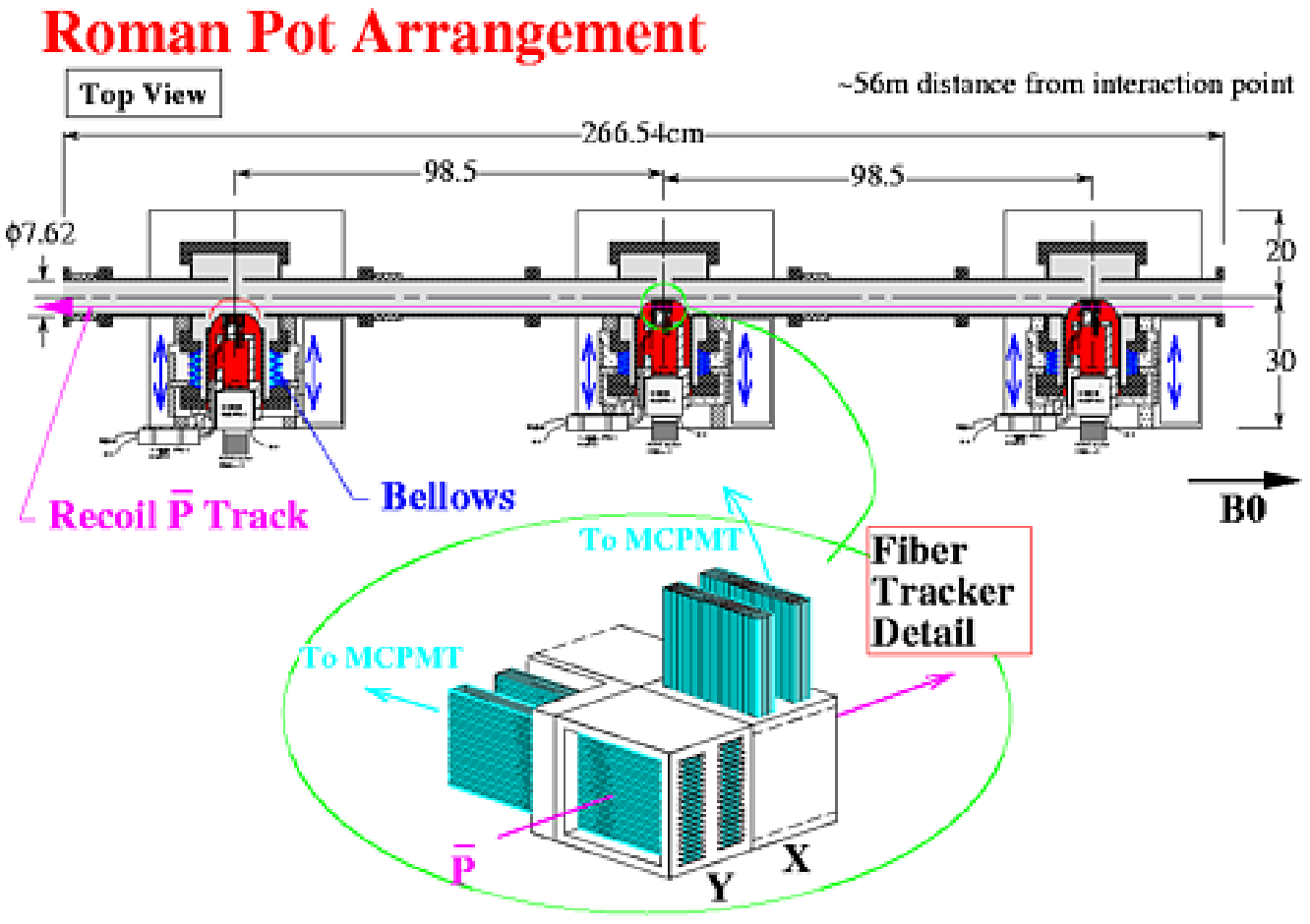}}
\caption{\label{fig:rps_detector} 
Schematic view of the Roman Pot fiber tracker detector.}
\end{minipage}
\end{figure}

\noindent
The RP is a fiber detector spectrometer with a 2-meter lever arm
located approximately 57~meters from the {\it Interaction Point} (IP) downstream 
of the anti-proton (Fig.~\ref{fig:rps_tevatron}).
The RP detector consists of three stations, approximately one meter apart from each
other (Fig.~\ref{fig:rps_detector}).
\noindent
Each station consists of one trigger counter and one 
80-channel scintillation fiber detector viewed by a Hamamatsu H5828 80-channel
Photo-Multiplier Tube (PMT). 
The fiber detector reads X (40 channels) and Y (40 channels) 
coordinates to identify the position of the tracks with a resolution 
of approximately $100~\mu$m.\\

\noindent
In preparation for Run~II, the detector was reinstalled as in Run~I.
In contrast, the read out electronics was completely redesigned to take into account the shorter 
(with respect to Run~I) 396~nsec~spacing between bunches.
Moreover, the beam polarity at the RP location was reversed with respect to Run~I, and the
anti-proton beam travels now closer to the RP detector. This will allow close to full
acceptance down to $\xi\sim0.03$ (in Run~I the region of full acceptance was $0.05<\xi <0.1$).\\
Lower values of $\xi$ are necessary to enhance the rate of diffractive events.
A coincidence signal of the three counters selects events with an outgoing
anti-proton at the RP location and is used as an input at Level 1 trigger. 
This coincidence signal is used together with other signatures in the CDF detector to
enhance the selection of diffractive events.

\subsection{Beam Shower Counters}
\begin{figure}[thb]
\begin{minipage}[]{1.\linewidth}
\epsfxsize=0.9\textwidth
\hspace*{0cm}
\centerline{\epsfbox[30 260 550 650]{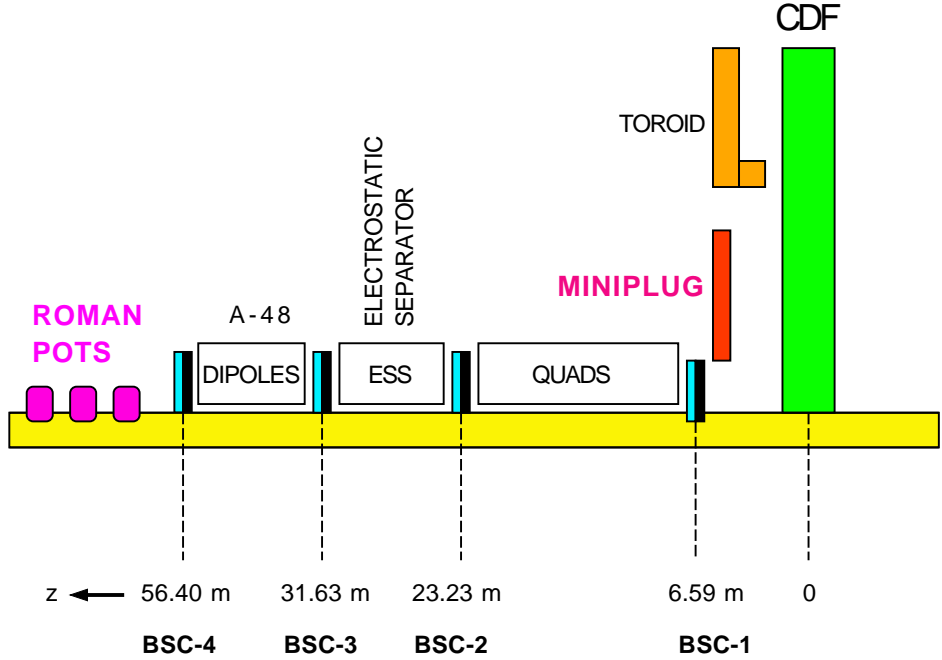}}
\caption{\label{fig:fd_detectors}
Location of the Forward Detectors along the $\overline{p}$ direction on the
West side of the CDF central detector (not to scale). On the East side, only
the first three BSC stations are installed, as there is no room for BSC-4.}
\end{minipage}
\end{figure}

\begin{figure}[thb]
\begin{minipage}[]{.45\linewidth}
\epsfxsize=1.1\textwidth
\vspace*{0.5cm}
\centerline{\epsfbox[10 120 750 620]{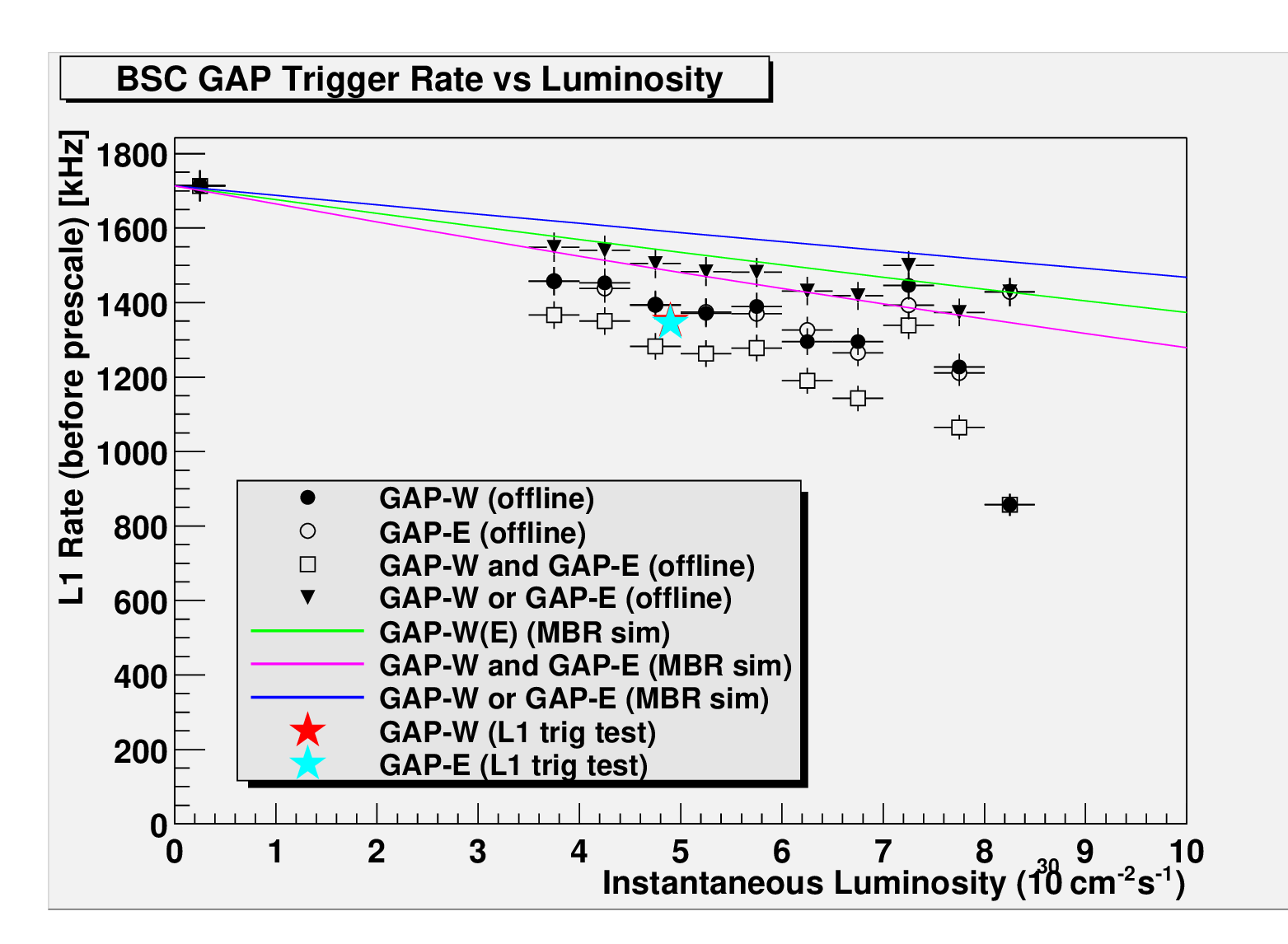}}
\vspace*{0.5cm}
\caption{\label{fig:bsc_triggers}
Level 1 {\it gap} trigger rates provided by the BSCs at different values of the luminosity.}
\end{minipage}
\hspace*{0.6cm}
\begin{minipage}[]{0.45\linewidth}
\epsfxsize=1.\textwidth
\centerline{\epsfbox[0 0 565 545]{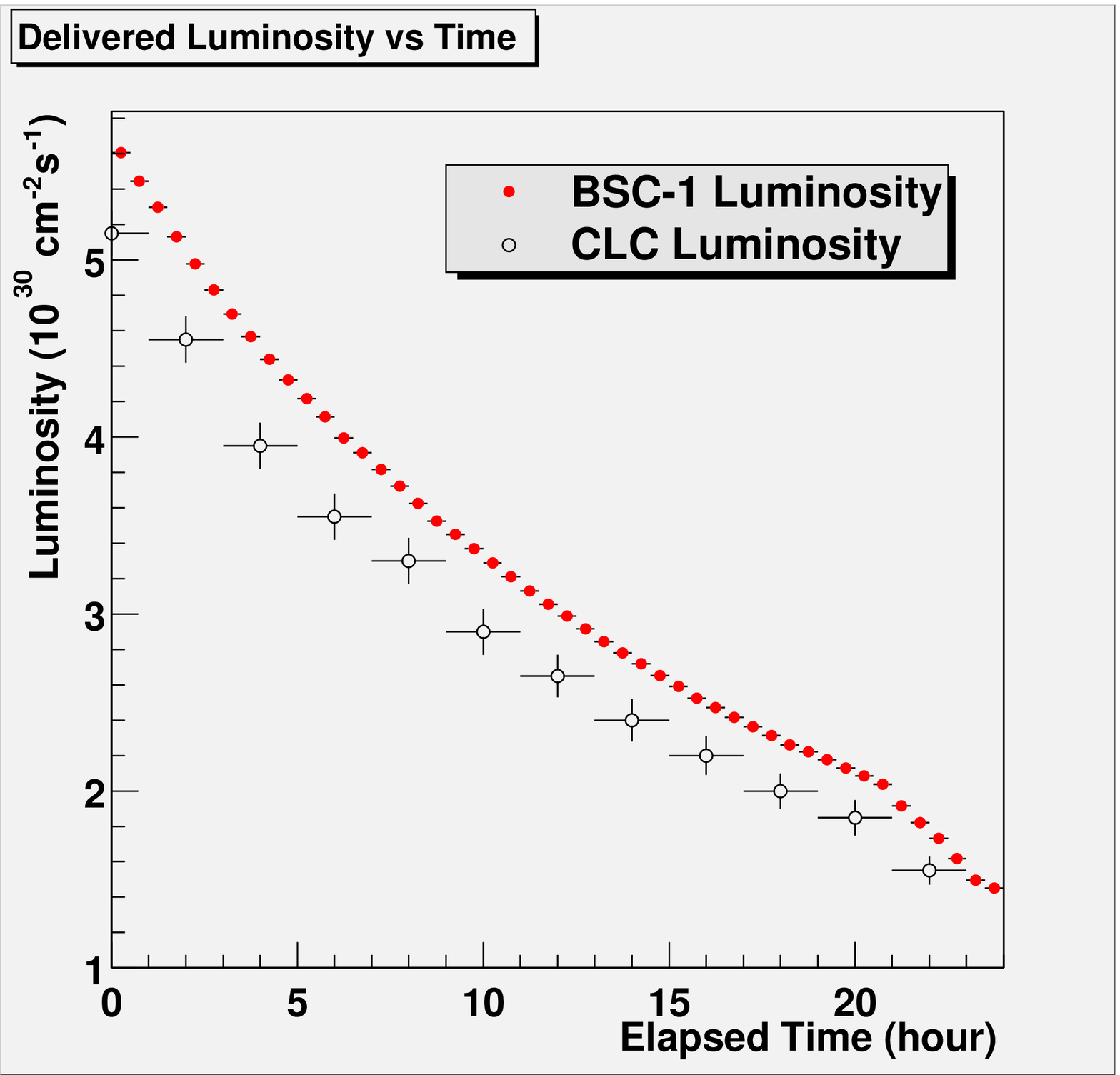}}
\caption{\label{fig:lum_bsc} 
The {\it East-West} coincidence rate from BSC-1 is compared with the
measurement provided by the \v Cerenkov Luminosity Counters.}
\end{minipage}
\end{figure}

\noindent
SD and DPE processes are characterized by forward rapidity gaps. 
An ideal rapidity gap tagger at the trigger level would only select these
events and can be accomplished with a set of scintillation counters around 
the beam-pipe at several locations along the $p$ and $\bar{p}$ directions 
covering the forward pseudorapidity region.
The BSCs can be used to identify diffractive events with a leading anti-proton, 
or also to veto non-diffractive minimum bias events. 
This is important for the inclusive RP and DPE triggers.\\
The BSCs detect particles traveling in either direction from the IP along 
and near the beam-pipe and cover the pseudorapidity region 5.5$<|\eta |<$7.5.
There are four BSC stations on the west side and three on the east side of
the IP.  All stations are located along the beam-pipe, at increasing distances
from the IP as one goes from BSC-1 to BSC-4 (Fig.~\ref{fig:fd_detectors}).  
BSC-1, 2 and 3 consist of two stations each, positioned symmetrically with 
respect to the IP, whereas BSC-4 consists of a single station on the west side.\\
Stations are made
of two scintillation counters, except for the BSC-1 
stations, which have four counters.  Since each counter is connected to its
own PMT, the entire system consists of 18 signal channels, 10 from the west 
and 8 from the east side.\\
The output signals from the PMTs on each side of the IP identify tracks traveling close to the beam-pipe.
BSCs separately provide East and West {\it gap} triggers at Level 1, 
when no signal is observed in either counter.
Trigger rates are satisfactorily within $\approx 10\%$ from Monte Carlo
expectations (Fig.~\ref{fig:bsc_triggers}) at different values of luminosity.
{\it Gap} triggers are then combined with other triggers at Level 2 and at Level 3
in order to provide a purer diffractive event sample.\\

\noindent
The BSC-1 counters can also be used to monitor Tevatron beam losses and
collision rates, as well as the {\it x-y} beam position.\\
The signals from BSC-1 are referred to as ``Out-of-Time'' for particles hitting
the counters $\sim 20$~nsec before interaction time (also due to beam losses),
and as ``In-Time'' for particles coming from a collision and hitting
the counters $\sim 20$~nsec after interaction time (collision rate and diffractive
physics measurements). 
The {\it East-West} coincidence trigger requires at least one BSC-1 counter on each side of the
interaction to fire within a 20~nsec window centered on the beam crossing time.
The luminosity $\cal{L}$ can be calculated from the collision rate using the formula:
$$ N[Hz]={\cal{L}}[Hz/mb]\cdot\sigma[mb]\cdot\epsilon , $$
where $\sigma$ is the $p\overline{p}$ cross section 
and $\epsilon$ is the acceptance of the counters.
The beam-beam cross section for the {$E\cdot W$} coincidence is calculated to be
$\sigma_{E\cdot W}=\sigma\times\epsilon$~=~25.6~mb.
The coincidence rate between BSC-1 East and BSC-1 West is 
shown in Figure~\ref{fig:lum_bsc} and is compared to the measurement by
the \v Cerenkov Luminosity Counter (CLC).\\
The BSC-1 counters are segmented in four quadrants in $\phi$, around the
beam-pipe. The $\phi$ segmentation allows real-time monitoring of
the {\it x-y} position of the beam. The information can be used by the 
Accelerator Beams' Division to steer the beam in the center of the CDF detector.

\subsection{MiniPlug Calorimeters}
\noindent
The program of hard diffraction and very forward physics for Run~II benefits from
two forward MP calorimeters in the pseudorapidity region
3.6$<|\eta|<$5.1 designed to measure the energy and lateral position of both
electromagnetic and hadronic showers. The MPs approximately double the pseudorapidity
region covered by the Plug calorimeters, which is 1.1$<|\eta|<$3.5. 
The MP and the Plug calorimeters can measure the width of the rapidity gap(s)
produced in diffractive processes and will allow extending the Run~I studies of
the diffractive structure function to much lower values of the fractions $\xi$,
where $\xi$ is the momentum of the proton carried by the Pomeron. 
The low $\xi$ values can be measured from the size of the rapidity gap
region using information from both BSCs and MPs.\\

\begin{figure}[thb]
\vspace*{0.5cm}
\begin{minipage}[]{.46\linewidth}
\epsfxsize=1.1\textwidth
\centerline{\epsfbox{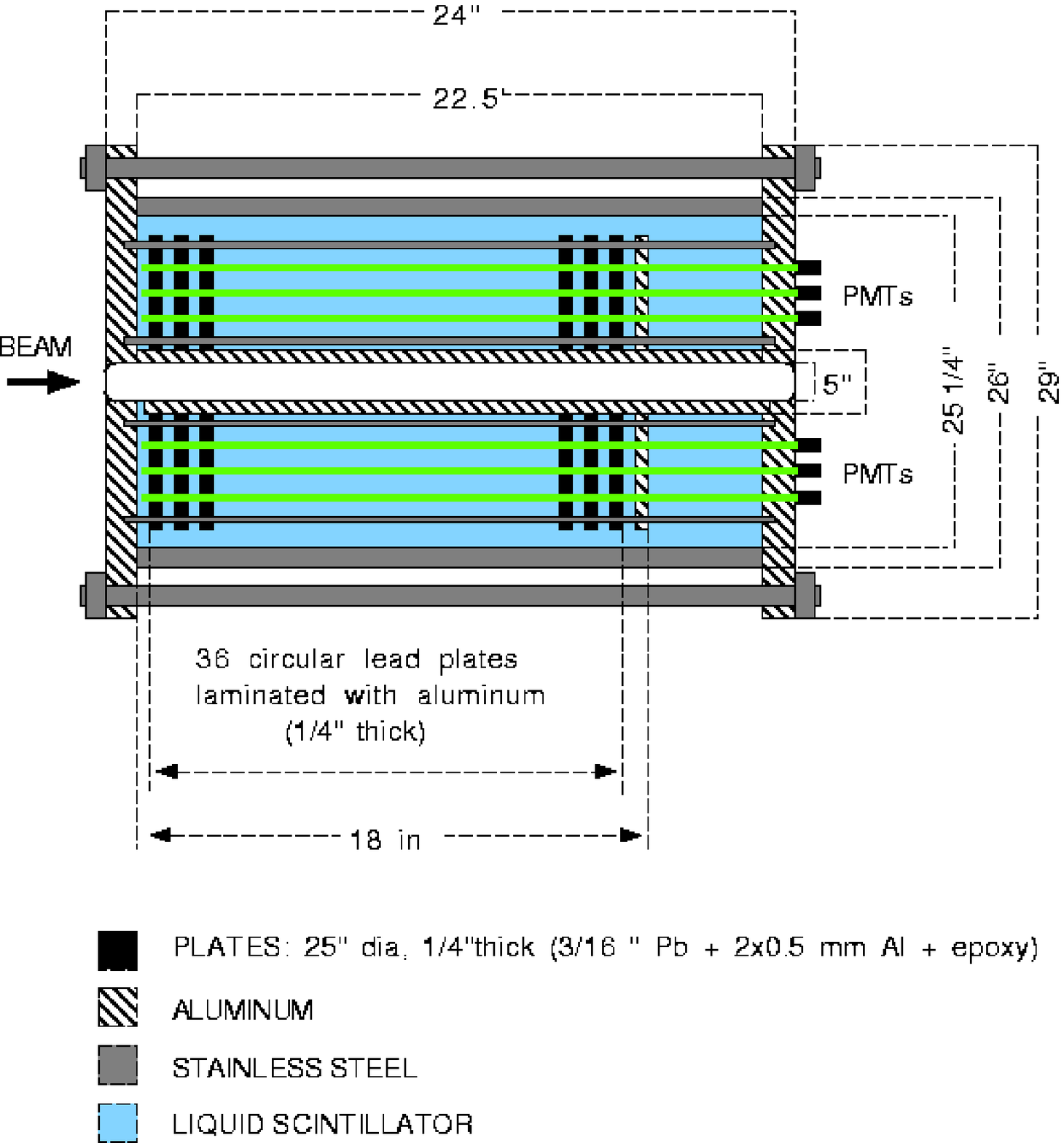}}
\vspace*{0.5cm}
\caption{\label{fig:mp_side} 
Schematic side view of a MiniPlug (not to scale).}
\end{minipage}
\hspace*{0.5cm}
\vspace*{1.cm}
\begin{minipage}[]{.46\linewidth}
\epsfxsize=1.1\textwidth
\centerline{\epsfbox{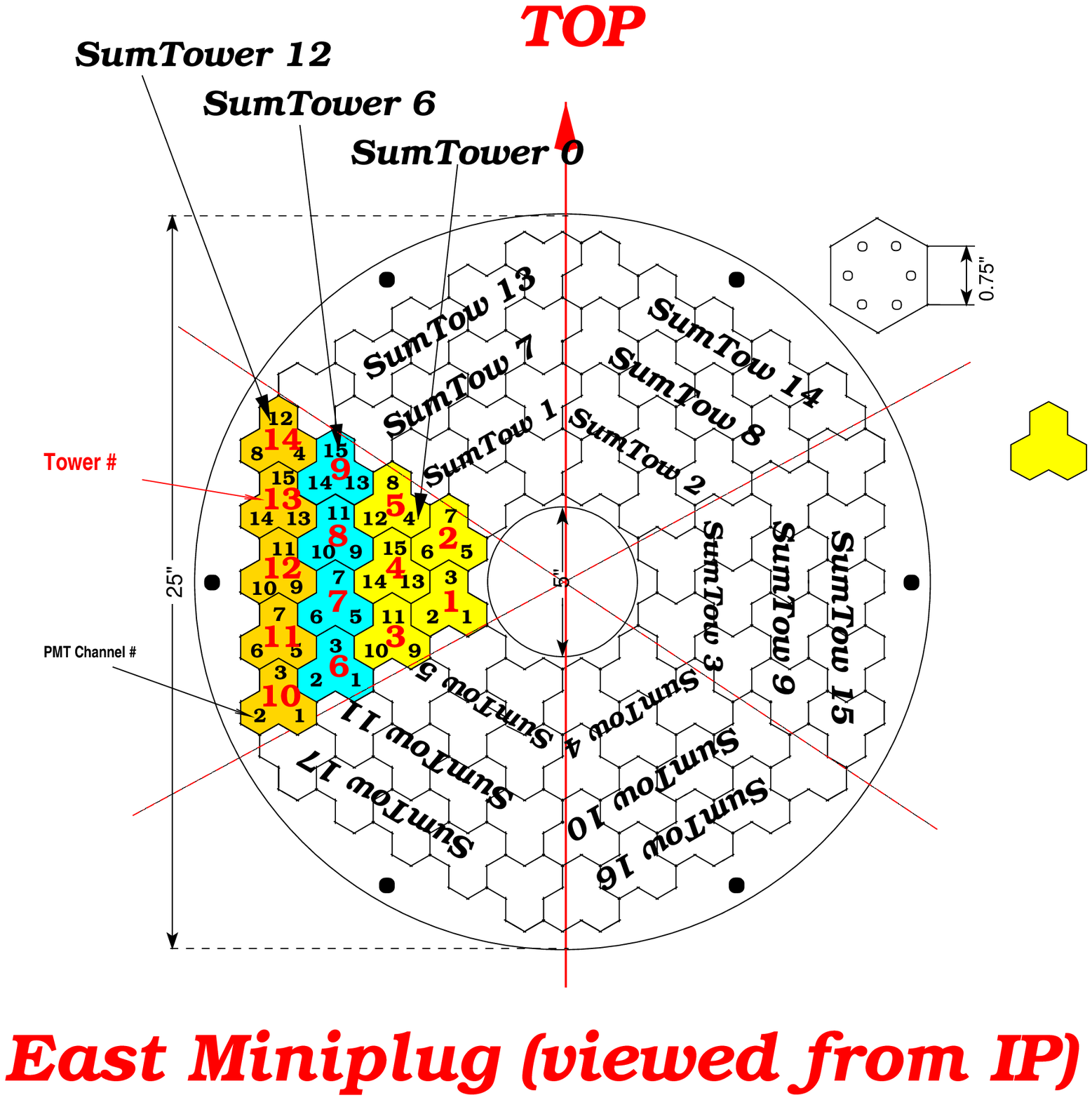}}
\vspace*{0.5cm}
\caption{\label{fig:mp_tower_geometry} 
Tower geometry of the East Miniplug calorimeter (viewed from the interaction point).}
\end{minipage}
\end{figure}

\noindent
The MPs consist of alternating layers of lead plates and liquid scintillator 
read out by wavelength shifting (WLS) fibers.
The WLS fibers are perpendicular to the lead plates and parallel to the 
proton/anti-proton beams in a geometry where {\it Towers} are formed by combining
the desired numbers of fibers
and read out by Multi-Channel PhotoMultipliers (MCPMTs).\\
Each MP is housed in a cylindrical steel barrel $\boldmath 26''$ in diameter and has a 
$\boldmath 5''$--hole concentric with the cylinder axis to accommodate the beam-pipe (Fig.~\ref{fig:mp_side}).
The active depth of each MP is 32 radiation lengths and 1.3 interaction lengths.\\
The tower geometry is organized in three concentric circles around the beam-pipe
(Fig.~\ref{fig:mp_tower_geometry})
with an {\it inner}, a {\it middle} and an {\it outer} ring. Each ring covers a different region
in pseudorapidity. This allows triggering on different $\eta$
regions, both for events with a {\it gap} region and for events with high-$E_T$ clusters.
\par \noindent
The design is based on a hexagon geometry. Uniformly distributed over each
plate, holes are conceptually grouped in hexagons and
each hexagon has six holes. 
A WLS fiber is inserted in each hole. The six fibers of one hexagon are grouped
together and are viewed by one MCPMT channel. 
The MCPMT outputs are added in groups of three to form 84 calorimeter towers.
The entire MCPMT can also be read out through the last dynode output,
indicated as {\it Sum-Tower} in Figure~\ref{fig:mp_tower_geometry},
and provides triggering information.
Each MP has a total of 18 Sum-Tower signals, divided in three rings.\\
An additional clear fiber
carries the light from a calibration LED to each MCPMT pixel.
The LED allows a first relative gain calibration to equalize the MCPMT from each ring.
It also allows periodical monitoring of the MCPMT response.
A cross-calibration between data and Monte Carlo expectations allows
an absolute energy calibration.\\

\begin{figure}[thb]
\vspace*{0.1cm}
\begin{minipage}[]{.45\linewidth}
\epsfxsize=1.1\textwidth
\centerline{\epsfbox[20 380 540 670]{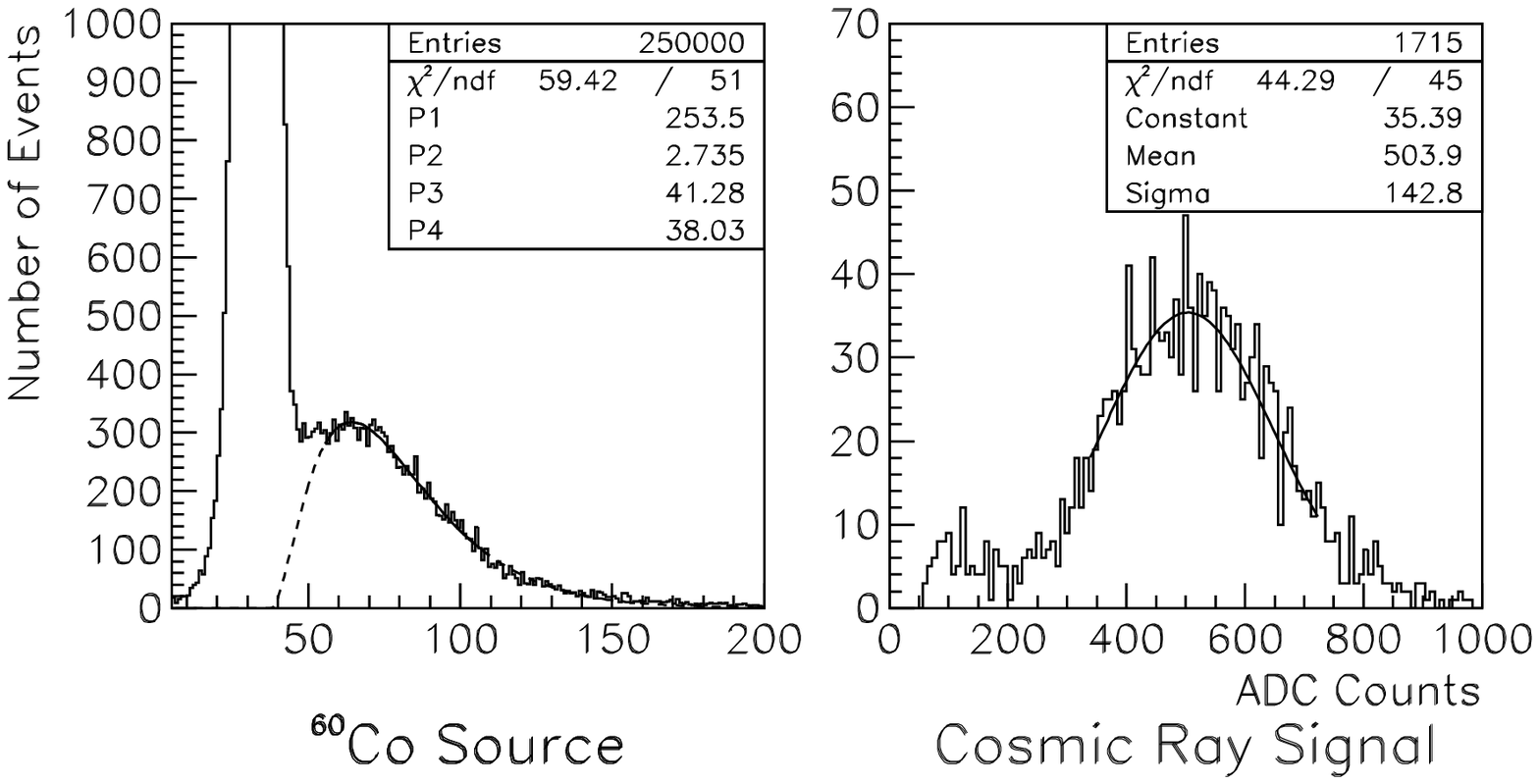}}
\vspace*{0.5cm}
\caption{\label{fig:cosmic_test} 
Cosmic ray test for {\it Tower} \#7 on East MiniPlug: 
$^{60}$Co source signals (left). The parameter {\it P3} corresponds to the
pulse height of a single photoelectron;
cosmic ray spectrum after an isolation requirement is fitted to a
Gaussian distribution (right).}
\end{minipage}
\hspace*{0.5cm}
\begin{minipage}[]{.45\linewidth}
\epsfxsize=1.1\textwidth
\centerline{\epsfbox[25 80 565 700]{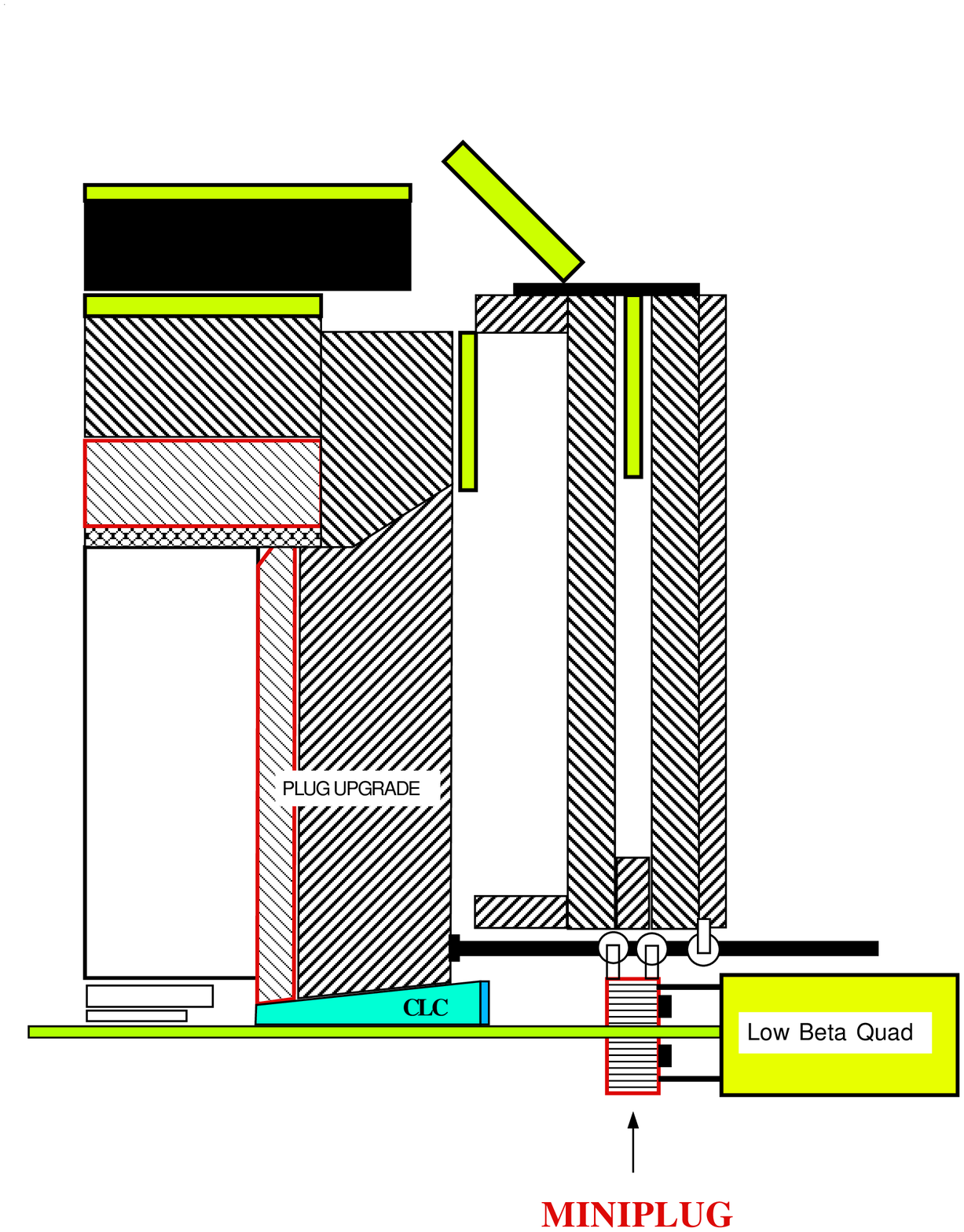}}
\caption{\label{fig:mp_at_cdf} 
Schematic drawing of a quarter view of the CDF detector showing a
MiniPlug calorimeter installed inside the toroids (not to scale).}
\end{minipage}
\end{figure}

\noindent
Cosmic ray muons were used to test one 60$^\circ$ wedge of the East MP.
In this test, the cosmic ray trigger fired on a 2-fold coincidence of the scintillation 
counter paddles located on top and on the bottom of the MP vessel,
placed with the towers pointing upward.
The outputs from Towers~\#6,~7,~8 and~9 and from Sum-Towers~\#2,~8 and~14 
(Fig.~\ref{fig:mp_tower_geometry}) were read out.
An energy isolation cut selects only those muons which went
through the entire length of the central Sum-Tower (\#8) and vetoed 
on the signals from the neighboring Sum-Towers (\#2 and 14).
The single photoelectron response for Sum-Tower \#8 is measured using 
a signal from a $^{60}$Co source randomly gated
(Fig.~\ref{fig:cosmic_test}).
The single tower response to a {\it minimum ionizing particle}
is estimated to be
approximately 120 photoelectrons, exceeding the design specification.\\
The MPs have been installed along the beam-pipe within the hole of 
the muon toroids at a distance of 5.8 meters from the center of the detector
(Fig.~\ref{fig:mp_at_cdf}).

\section{Diffractive Triggers}
\noindent
The triggers for diffractive physics are needed to enhance the selection of
the diffractive events.
All of the Forward Detectors provide triggering information to the {\it Level~1}
trigger selection, which is then combined together with other
physical entities from the rest of the CDF central detector such as, for example, large transverse
energy {\it jets}. The events selected by the diffractive triggers are then written on disk 
at a rate of approximately 2~Hz.\\

\section{Conclusions and Prospects}
\noindent
The program for diffractive physics during Run~II at the Tevatron includes
studies of hard diffraction and of double Pomeron exchange.
Some of these topics and desired goals for the immediate future 
have been discussed here.\\
In addition to other sub-detector parts, the CDF experiment is preparing
for the next few years of data-taking with a detector that has been 
exceptionally improved in many areas.
The Forward Detector upgrade project is essential for the program 
of diffractive physics studies.
The detectors have been installed, the commissioning is well underway,
and some of the results have been presented here.
The Forward Detectors are getting ready to take good quality physics data as
the Tevatron is reaching its design goals for luminosity.\\
These data will hopefully answer some of the questions addressed here 
and allow a better understanding of phenomena in 
hard diffraction and very forward physics at the Tevatron in \ppbar collisions.\\
In conclusion, Run~II is finally becoming a reality after many years of
detailed studies, accurate designs, promising improvements and intense upgrading.

\section{Acknowledgments}
\noindent
My warm thanks to the organizers of this conference for a wonderful workshop, and to all 
the people who strenuously contributed to this multi-year project.


\end{document}